\documentclass[conference]{IEEEtran}
\IEEEoverridecommandlockouts
\usepackage{cite}
\usepackage{comment}
\usepackage{amsmath,amssymb,amsfonts}

\usepackage{algpseudocode}
\usepackage{graphicx}
\usepackage{subfigure}
\usepackage{float}
\usepackage{subfig}
\usepackage[font=footnotesize,labelfont=rm]{caption}
\usepackage{textcomp}
\usepackage{xcolor}
\usepackage[linesnumbered,ruled,vlined]{algorithm2e}
\SetKwInput{KwInput}{Input}                
\SetKwInput{KwOutput}{Output}              

\def\BibTeX{{\rm B\kern-.05em{\sc i\kern-.025em b}\kern-.08em
    T\kern-.1667em\lower.7ex\hbox{E}\kern-.125emX}}

\begin{document}

\title{Ultra-Reliable Low-Latency Communication for Aerial Vehicles via Multi-Connectivity}

\author{\IEEEauthorblockN{Fateme Salehi\IEEEauthorrefmark{1}\IEEEauthorrefmark{2}, Mustafa Ozger\IEEEauthorrefmark{1}, Naaser Neda\IEEEauthorrefmark{2}, Cicek Cavdar\IEEEauthorrefmark{1}}
\IEEEauthorblockA{\IEEEauthorrefmark{1}School of Electrical Engineering and Computer Science,  KTH Royal Institute of Technology, Sweden}

\IEEEauthorblockA{\IEEEauthorrefmark{2}Faculty of Electrical and Computer Engineering, University of Birjand, Iran}
Email: \{fatemes, ozger, cavdar\}@kth.se, nneda@birjand.ac.ir
}

\maketitle

\begin{abstract}
Aerial vehicles (AVs) such as electric vertical take-off and landing (eVTOL) make aerial passenger transportation a reality in urban environments. However, their communication connectivity is still under research  to realize their safe and full-scale operation, which requires stringent end-to-end (E2E) reliability and delay. In this paper, we evaluate  reliability and delay for the downlink communication of AVs, i.e., remote piloting, control/telemetry traffic of AVs. We investigate direct air-to-ground (DA2G) and air-to-air (A2A) communication technologies, along with high altitude platforms (HAPs) to explore the conditions of how multi-connectivity (MC) options satisfy the demanding E2E connectivity requirements under backhaul link bottleneck. Our considered use case is ultra-reliable low-latency communication (URLLC) under the finite blocklength (FBL) regime due to the nature of downlink control communication to AVs. In our numerical study, we find that providing requirements by single connectivity to AVs is very challenging due to the line-of-sight (LoS) interference and reduced gains of downtilt ground base station (BS) antenna. We also find that even with very efficient interference mitigation, existing cellular networks designed for terrestrial users are not capable of meeting the URLLC requirements calling for MC solutions.

\end{abstract}


\section{Introduction}\label{sec.intro}

5G and beyond systems are expected to support ultra-reliable low-latency communications (URLLC) to provide connectivity to mission-critical services in which uninterrupted and robust data exchange is vital \cite{URLLC}. One critical emerging scenario for URLLC is remote piloting of aerial vehicles (AVs) where connectivity has a key role to ensure safe operations. AVs such as electric vertical take-off and landing (eVTOL) vehicles enable passengers to be transported over several tens of kilometers at low altitudes as an extension of urban transportation systems \cite{FACOM}. They have different mission and flight characteristics with diverse quality of service (QoS) requirements such as data rate, end-to-end (E2E) latency and communication reliability. In command and control (C2) links, short-packets are transmitted to navigate the AVs remotely via URLLC.   

A scenario for beyond visual line-of-sight (BVLoS) operation for remote piloting of unmanned aerial vehicles (UAVs) is studied in \cite{BVLOS}, which utilizes different technologies such as mobile edge computing and augmented reality. Multi-hop air-to-air (A2A) and direct air-to-ground (DA2G) communication is considered in \cite{aircraft} to increase the performance of DA2G communication without considering the reliability performance. The authors of \cite{URLLC_UAV} exploit the macro-diversity gain of the distributed multi-antenna systems and the array gain of the centralized multi-antenna systems. They maximize the available range of the C2 communication links between UAVs and a ground base station (BS) by optimizing the altitude of UAVs, the duration of the uplink and downlink phases, and the antenna configuration.

\vspace{-1mm}
One of the methodologies to provide URLLC services is to utilize multi-connectivity (MC). Introducing link or radio access technology (RAT) diversity can improve latency and reliability performance. In this regard, the authors of \cite{drone_dc} and \cite{dual_MNO} conduct field measurements with a UAV to evaluate the improvements in reliability and latency over multiple mobile network operators (MNOs).  
In \cite{maritime}, a combination of a public and dedicated cellular network with multipath transmission control protocol (MPTCP) is proposed for maritime search and rescue missions of UAVs. The results show that the multi-link protocol increases the range and improves the data rate performance.
The authors of \cite{multipath} aim to provide robust bandwidth allocation as well as high dynamic system stability. To this end, they present an analytic modeling with a satellite link and WiFi access points using MPTCP to control a swarm of UAVs. The authors of \cite{3fold_redundant} present field measurements of triple-redundant multi-link architecture employing cellular, WiFi and LoRa for the UAV-ground station connectivity. Their redundancy design employs the cellular network as the primary link and the other two as fallback links when there is no cellular coverage. None of the previous studies consider E2E paths for C2 communications considering backhaul links and network architectures.

\vspace{-1mm}
Different from prior works in \cite{BVLOS,aircraft,URLLC_UAV,drone_dc,dual_MNO,maritime,multipath,3fold_redundant}, the main contribution of this paper is the analysis of E2E delay and reliability of C2 communications considering backhaul links and network architectures with practical antenna configurations and finite blocklength (FBL) regime. 
As concepts of eVTOLs are of recent venture, to the best of our knowledge, the literature has not yet covered the connectivity needs and potential solutions for the C2 communication links of these AVs. In particular, we consider a heterogeneous network of ground BSs, relay AVs, and a high altitude platform (HAP) with unreliable backhaul link to provide connectivity for AVs. We take into account ground BSs with downtilt antennas and HAP with realistic antenna patterns. The automatic repeat-request (ARQ) mechanism and frequency diversity is employed to improve reliability of radio links. Next, we investigate how multi-path connectivity by  multi-RAT can guarantee URLLC requirements of remote piloting of AVs. 

The paper is organized as follows. Section \ref{sec.sys} presents the system model. 
Section \ref{sec.chmodel} presents the channel models of communication links, with antenna radiation pattern of ground BSs and HAP.
Section \ref{sec.anal} presents E2E delay and reliability analysis including different RATs. Section \ref{sec.res} discusses the numerical results and finds the required parallel links to enable remote piloting operation in different communication distances. Section \ref{sec.con} concludes the paper.

\vspace{-1mm}
\section{System Model}\label{sec.sys}
In this section, we introduce the scenario under study with its requirements and the methodology for providing them.
\subsection{Remote Piloting of AVs and QoS Requirements}

BVLoS remote piloting of an AV requires a communication path between the remote pilot and the AV. We analyze the reliability and delay of this E2E path considering not only wireless single hop links but also backhaul links connecting BSs to the core network. Based on \cite{FACOM}, control/telemetry traffic for remote piloting operations of eVTOLs requires a data rate about $0.25-1$ Mbps, E2E latency less than $10-150$ ms, and the minimum communication reliability $99.999\%$.

\vspace{-1mm}
\subsection{Multi-Connectivity}
\begin{figure}[b!]
\vspace{-3mm}
    \centering
            \includegraphics[width=0.98\columnwidth]{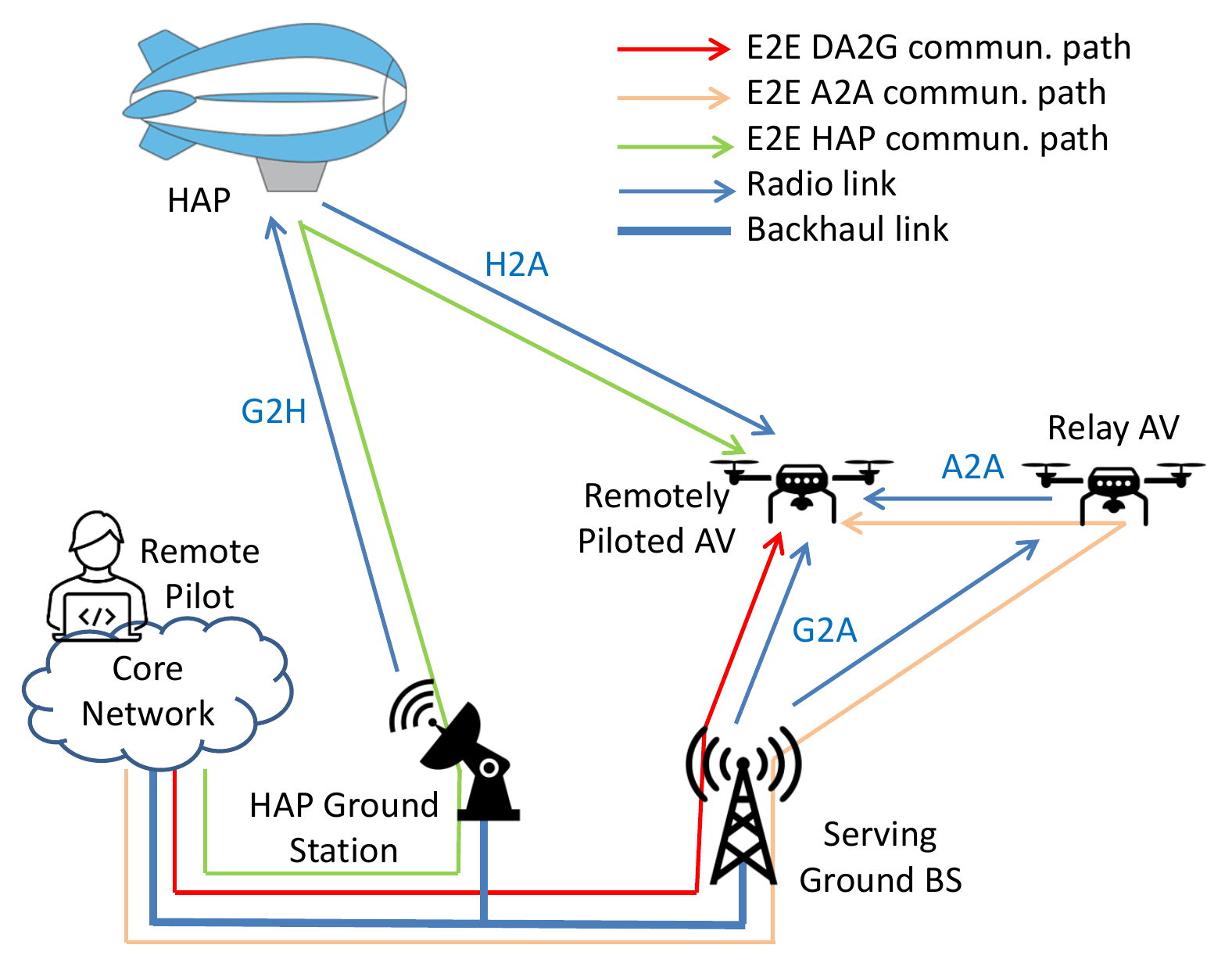}
            \caption{Illustration of multi-RAT and E2E communication paths.}
            \label{fig.sys}
\end{figure}

MC using multiple communication paths simultaneously is a promising approach to fulfill strict requirements of AVs' remote piloting. The considered system model consists of an integration of multiple RATs including DA2G, A2A and HAP communication. The E2E path of each RAT is illustrated in Fig. \ref{fig.sys}, a directive path starting with the core network, traversing the backhaul link and the radio link/links to reach the destination AV, which is the AV that remote pilot wants to navigate. The radio communication links consist of ground BS-to-AV (G2A), ground station-to-HAP (G2H), and AV/HAP-to-AV (A2A/H2A). In Fig. \ref{fig.sys}, three different E2E paths are shown. The red line illustrates ``DA2G path", which includes the backhaul link to the ground BS and G2A link. ``A2A path", illustrated with orange line is defined as the path consisting of backhaul, G2A and A2A links. The green line illustrates the ``HAP path" defined as the path consisting of backhaul link to the ground station, G2H and H2A links.   

\vspace{-1mm}

\subsection{Transmission and Combining Strategy}
We consider packet cloning for transmitting the message from the remote pilot to the AV over independent links. In this approach, the source sends copies of the message through each of the available links \cite{interface_diversity}. The combining scheme is joint decoding, where each link is decoded individually. Hence, the E2E error probability of $N$ parallel transmission paths is 
\vspace{-1mm}
\begin{equation}
\label{eq.error.tot}
\vspace{-1mm}
    \mathcal{E}_{\rm{E2E}}=\prod_{i=1}^{N} \mathcal{E}_{\rm{E2E}}^{i},
\end{equation}
where $\mathcal{E}_{\rm{E2E}}^{i}$ is the error probability of the $i$th path, and $i\in \{\rm{g},\rm{a},\rm{h}\}$ refers to different RATs including DA2G, A2A, and HAP communications, respectively. It potentially reduces the delay, since only the packet that arrives earlier and is decoded correctly needs to be considered. Hence, the E2E delay of multi-RAT using the cloning scheme is calculated as \cite{interface_diversity}
\vspace{-1mm}
\begin{equation}
\label{eq.delay.tot}
    \mathcal{D}_{\rm{E2E}}=\min _{i} \left\{\mathcal{D}_{\rm{E2E}}^{i}\right\},
\end{equation}
where $\mathcal{D}_{\rm{E2E}}^{i}$ is the E2E delay of the $i$th path.

MC can provide URLLC services without intervention in the physical layer design. Although it may not be resource-efficient by using packet cloning strategy, the efficiency can be improved by other transmission schemes.

\section{Channel Models of Communication Links and Antennas Radiation Patterns}\label{sec.chmodel}
To model a realistic propagation channel, we consider both large-scale fading and small-scale fading.
\subsection{Large-Scale Fading}
\subsubsection{Path Loss of G2A Link}
We consider that the G2A link experiences LoS propagation with a probability of $P_\mathrm{LoS}$, which is given by \cite{U2U_comm}
\begin{equation}
  P_\mathrm{LoS}=\prod_{j=0}^{k}\left[1-\exp \left(-\frac{\left[h_{\mathrm{g}}-\frac{(j+0.5)\left(h_{\mathrm{g}}-h_{\mathrm{a}}\right)}{k+1}\right]^{2}}{2 \mathrm{q}_{3}^{2}}\right)\right],  
\end{equation}
where $k=\left\lfloor\frac{r_\mathrm{ga} \sqrt{\mathrm{q}_{1} \mathrm{q}_{2}}}{1000}-1\right\rfloor$ and $r_\mathrm{ga}$ is the 2D distance between the ground BS and the AV, while $\{\mathrm{q}_{1}, \mathrm{q}_{2}, \mathrm{q}_{3}\}$ are environment-dependent parameters set to $\{0.3, 500, 20\}$ to model an urban scenario \cite{U2U_comm}. Moreover, $h_{\mathrm{g}}$ and $h_{\mathrm{a}}$ are the height of the ground BS and altitude of the AV, respectively.

Thus, the average path loss of the G2A link is derived as
\begin{equation}
\label{eq.PL_ga}
  \mathit{PL}_\mathrm{ga}=P_\mathrm{LoS} \times \mathit{PL}^{\mathrm{LoS}}_\mathrm{ga}+(1-P_\mathrm{LoS}) \times \mathit{PL}^{\mathrm{NLoS}}_\mathrm{ga},
\end{equation}
where $\mathit{PL}^{\mathrm{LoS}}_\mathrm{ga}$ and $\mathit{PL}^{\mathrm{NLoS}}_\mathrm{ga}$ are the path losses of the G2A channel under LoS and NLoS conditions, respectively. Based on the urban macro cells (UMa) scenario, $\mathit{PL}^{\mathrm{LoS}}_\mathrm{ga}$ and $\mathit{PL}^{\mathrm{NLoS}}_\mathrm{ga}$ are calculated as follows \cite{3GPP36.777}
\begin{equation}
    \mathit{PL}^{\mathrm{LoS}}_\mathrm{ga}(\mathrm{dB})=28+22\log_{10} \left(d_\mathrm{ga}\right) + 20\log_{10} \left( f_{\rm{c}} \right),
\end{equation}
\begin{equation}
\begin{split}
    \mathit{PL}^{\mathrm{NLoS}}_\mathrm{ga}(\mathrm{dB})&=-17.5+\left(46 - 7\log_{10} \left( h_{\rm{a}} \right)\right)\log_{10} \left(d_\mathrm{ga}\right)\\
    &+ 20\log_{10} \left( 40\pi f_{\rm{c}}/3 \right),    
\end{split}
\end{equation}
where $d_\mathrm{ga}$ is the 3D distance between the ground BS and the AV in meter, and $f_{\rm{c}}$ is the carrier frequency in GHz.

\subsubsection{Path Loss of A2A/H2A Link}
For A2A and H2A links, the free space path loss (FSPL) channel model is used \cite{3GPP38.811}:
\begin{equation}
\begin{split}
     \mathit{PL}_{\rm{xy}}{(\rm{dB})} &= \mathit{FSPL}(d_{\rm{xy}},f_{\rm{c}})\\ &= 32.45 + 20 \log_{10} \left( d_{\rm{xy}} \right) + 20\log_{10} \left( f_{\rm{c}} \right),    
\end{split}
\end{equation}
where $\rm{xy}\in\{\rm{aa},\rm{ha}\}$ represents the A2A and the H2A link, respectively. $d_{\rm{xy}}$ is the 3D distance between nodes $\mathrm{x}$ and $\mathrm{y}$ in meter, and $f_{\rm{c}}$ is the carrier frequency in GHz.

\subsubsection{Path Loss of G2H Link}
The path loss of the G2H link can be considered as the basic path loss model which accounts for the signal's free space path loss, shadow fading (SF), and clutter loss (CL) \cite{3GPP38.811} 
\begin{equation}
    \mathit{PL}_{\rm{gh}}{(\rm{dB})} = \mathit{FSPL}(d_{\rm{gh}},f_{\rm{c}}) + \mathit{SF} + \mathit{CL}.
\end{equation}
Shadow fading is modeled by a log-normal distribution, i.e., $\mathit{SF}\sim N(0,\sigma_{\mathit{SF}}^2)$. Clutter loss based on \cite[Table 6.6.2-1]{3GPP38.811} depends on the elevation angle between the ground station and HAP, the carrier frequency, and the environment. 

\subsection{Small-Scale Fading}
Due to the LoS path for all the mentioned links, small-scale channel fading between nodes $\mathrm{x}$ and $\mathrm{y}$, i.e., $\omega_\mathrm{xy}$, can be taken into account as the Rician model, where $\rm{xy} \in \{\rm{ga},\rm{aa},\rm{gh},\rm{ha}\}$.
\begin{equation}
f_{\Omega}(\omega_\mathrm{xy})=\frac{\omega_\mathrm{xy}}{\sigma_\mathrm{xy}^{2}} \exp \left(\frac{-\omega_\mathrm{xy}^{2}-\rho_\mathrm{xy}^{2}}{2 \sigma_\mathrm{xy}^{2}}\right) I_{0}\left(\frac{\omega_\mathrm{xy} \rho_\mathrm{xy}}{\sigma_\mathrm{xy}^{2}}\right),
\end{equation}
reflects the PDF of Rice distribution with $\omega_\mathrm{xy} \geq 0$, and $\rho_\mathrm{xy}$ and $\sigma_\mathrm{xy}$ the strength of the LoS and the NLoS paths, respectively. $I_{0}(.)$ denotes the modified Bessel function of the first kind and zero order. The Rice factor of X2Y link is defined as
 \begin{equation}
     K_\mathrm{xy}({\rm{dB}})=10\log_{10}\left(\frac{\rho_\mathrm{xy}^{2}}{2 \sigma_\mathrm{xy}^{2}}\right),
 \end{equation}
which increases directly with different parameters such as altitude, elevation angle, and carrier frequency. The elevation angle plays a dominant role among the other factors \cite{Rice_factor}.

\subsection{Antenna Gain}
We assume that all AVs are equipped with a single omnidirectional antenna with unitary gain. However, we consider a realistic antenna radiation pattern for the ground BSs and the HAP, which is given as follows. 
\subsubsection{Ground BS Antenna Pattern}
We assume that the ground BSs are equipped with a vertical, $N$-element uniform linear array (ULA), where each element is omnidirectional in azimuth with maximum gain of $g_{\rm{E}}^{\max}$ and directivity as a function of the zenith angle $\phi$ \cite{U2U_comm}:
\begin{equation}
g_{\rm{E}}(\phi)=g_{\rm{E}}^{\max } \sin ^{2} \phi.
\end{equation}
We assume that there is a half-wavelength spacing between the adjacent antenna elements. With a fixed downtilt angle $\phi_{\rm{t}}$, the array factor of the ULA is given by \cite{U2U_comm}
\begin{equation}
g_{\rm{A}}(\phi)=\frac{\sin ^{2}\left(N \pi\left(\cos \phi-\cos \phi_{\mathrm{t}}\right) / 2\right)}{N \sin ^{2}\left(\pi\left(\cos \phi-\cos \phi_{\mathrm{t}}\right) / 2\right)}.
\end{equation}
The total ground BS's antenna gain in linear scale is 
\begin{equation}
  g_{\rm{g}}(\phi)=g_{\rm{E}}(\phi)\times g_{\rm{A}}(\phi).
\end{equation}

\subsubsection{HAP Antenna Pattern}
The following normalized  antenna gain pattern of one beam, corresponding to a typical reflector antenna with a circular aperture with radius of 10 wavelengths, is considered \cite{3GPP38.811}
\begin{equation}
g_{\rm{h}}(\theta)=\left\{\begin{array}{ll}
1, & \text { for } \theta=0, \\
4\left|\frac{J_{1}(20\pi \sin \theta)}{20\pi \sin \theta}\right|^{2}, & \text { for } 0<|\theta| \leq 90^{\circ},
\end{array}\right.
\end{equation}
where $\theta$ is the angle with respect to antenna boresight, and $J_1(.)$ is the Bessel function of the first kind and first order. It is assumed that each cell is served by one main beam \cite{HAP}.

\subsection{SINR Calculation}
One may obtain the channel coefficient between any two nodes $\mathrm{x}$ and $\mathrm{y}$ as
\begin{equation}
    h_\mathrm{xy} = (\frac{g_{\rm{xy}}}{\mathit{PL}_\mathrm{xy}})^{1/2}\omega_\mathrm{xy},
\end{equation}
where $g_{\rm{xy}}$ is the total antenna gain between nodes $\mathrm{x}$ and $\mathrm{y}$ given by the product of their respective antenna gains. Finally, the SINR of X2Y link with bandwidth $B^{\rm{xy}}$, $\rm{xy} \in \{\rm{ga},\rm{aa}\}$, is calculated as follows
\begin{equation}
    \gamma^{\rm{xy}} = \frac{p_{\rm{x}}|h_\mathrm{xy}|^2}{P_{\rm{interf}}\sum\limits_{i \in \mathcal{N}_i} {p_{\mathrm{x}_i}} |{h_{\mathrm{x}_i\mathrm{y}}}{|^2}+B^{\rm{xy}}N_0},
\end{equation}
where $p_{\rm{x}}$ is the transmit power of node $\mathrm{x}$, $P_{\rm{interf}}$ is the probability of interference, and $N_0$ is the noise spectral density. $\mathcal{N}_i$ is the set of interfering nodes and, $h_{\mathrm{x}_i\mathrm{y}}$ indicates the channel coefficient between the interfering node $\mathrm{x}_i$ and node $\mathrm{y}$. We assume that the G2H link is interference free, while the interference on H2A links is due to the side lobes of HAP's antenna overlapping the main lobes \cite{HAP}.

\section{Reliability and Latency Analysis}\label{sec.anal}
\subsection{Preliminaries}
The achievable data rate of the X2Y link, $R^{\rm{xy}}$, with FBL coding and an acceptable Block Error Rate (BLER) $\varepsilon_{\rm{t}}^{\rm{xy}}$, $\rm{xy} \in \{\rm{ga},\rm{aa},\rm{gh},\rm{ha}\}$, has an approximation as \cite{joint_UlDl}
\begin{equation}
    \label{eq.rate}
    R^{\rm{xy}}\approx B^{\rm{xy}}\left(\log_2(1+\gamma^{\rm{xy}})-\sqrt{\frac{V^{\rm{xy}}}{B^{\rm{xy}}D_{\rm{t}}^{\rm{xy}}}}\frac{Q^{-1}(\varepsilon_{\rm{t}}^{\rm{xy}})}{\ln2}\right)\text{bits/s},
\end{equation}
where $D_{\rm{t}}^{\rm{xy}}$ is the transmission delay of the X2Y link. Moreover, $Q^{-1}(\cdot)$ refers to the inverse Gaussian Q-function and $V^{\rm{xy}}=1-(1+\gamma^{\rm{xy}})^{-2}$  is the channel dispersion. 

Based on \eqref{eq.rate}, the decoding error probability is given by
\begin{equation}
    \label{eq.tr.error}
    \varepsilon_{\rm{t}}^{\rm{xy}}\approx Q\left(\frac{\left(B^{\rm{xy}}\log_2(1+\gamma^{\rm{xy}})-R^{\rm{xy}}\right)\ln2}{\sqrt{B^{\rm{xy}}V^{\rm{xy}}/D_{\rm{t}}^{\rm{xy}}}}\right).
\end{equation}
When transmitting a packet that contains $b$ bits over the allocated channel, the decoding error probability can be obtained by substituting $R^{\rm{xy}}=b/D_{\rm{t}}^{\rm{xy}}$ into \eqref{eq.tr.error}.

The above expressions are for AWGN channels which contain no fading. Hence, we can assume our channel as a quasi-static flat fading channel such that at each realization, its characteristics remain the same. In such a case, the expected value of $\varepsilon_{\rm{t}}^{\rm{xy}}$ is calculated  over $\gamma^{\rm{xy}}$, i.e., $\overline{\varepsilon}_{\rm{t}}^{\rm{xy}}$, since SINR changes with different channel realizations due to fading.
 
By adopting ARQ scheme, the packet is retransmitted until it is received correctly, and we assume that there is a reliable feedback from the AV to the transmitter as in \cite{ARQ}. Hence, the average transmission delay of the X2Y link is calculated as
\begin{equation}
    {\overline{D}_{\rm{t}}^{\rm{xy}}}=\frac{{D_{\rm{t}}^{\rm{xy}}}}{1-\overline{\varepsilon}_{\rm{t}}^{\rm{xy}}}.
\end{equation}

Denote the packet dropping probability due to queueing delay violation as
\begin{equation}
     \varepsilon_{\rm{q}}^{\rm{x}}=\operatorname{Pr}\left \{D_{\rm{q}}^{\rm{x}}>D_{\text{max}}^{\rm{q}} \right \},
\end{equation}
where $D_{\rm{q}}^{\rm{x}}$ is the queue delay of node $\rm{x}$, and $\rm{x} \in \{\rm{g},\rm{a},\rm{h}\}$ refers to ground BS, AV, and HAP, respectively. 
The packet arrival process of node $\rm{x}$ can be modeled as a Poisson process with the average arrival rate of $\lambda_{\rm{x}}$ packets/s \cite{joint_UlDl}. The effective bandwidth of the arrival process, which is the minimal constant packet service rate required to satisfy the queueing delay requirement $(D_{\text{max}}^{\rm{q}},\varepsilon_{\rm{q}}^{\rm{x}})$ can be expressed as follows \cite{joint_UlDl}
\begin{equation}
E_{\mathrm{BW}}^{\rm{x}}=\frac{ \ln \left(1 / \varepsilon_{\mathrm{q}}^{\rm{x}}\right)}{D_{\mathrm{q}}^{\rm{x}} \ln \left[\frac{ \ln \left(1 / \varepsilon_{\mathrm{q}}^{\rm{x}}\right)}{\lambda_{\rm{x}} D_{\mathrm{q}}^{\rm{x}}}+1\right]} \text{ packets/s. }
\end{equation}

\subsection{E2E Delay and Packet Loss Probability}
\subsubsection{E2E Path through DA2G Communication}
The E2E delay of DA2G path consists of delay due to backhaul link, $D_{\mathrm{b}}$, queue delay in the ground BS, $D_{\mathrm{q}}^{\rm{g}}$, and the average transmission delay of the G2A link, $\overline{D}_{\mathrm{t}}^{\rm{ga}}$. Hence, the E2E delay requirement can be satisfied with the following constraint
\begin{equation}
\label{eq.delayG2A}
\vspace{-2mm}
     D_{\mathrm{b}} + D_{\mathrm{q}}^{\rm{g}} +  \overline{D}_{\mathrm{t}}^{\rm{ga}} \leq D_{\mathrm{max}},
\end{equation}
where $D_{\mathrm{max}}$ is the required delay threshold. By deploying fiber optic backhaul links, we assume that the backhaul delay for remote piloting is around 1 ms\footnote{This value of backhaul delay corresponds to the propagation delay in a path with a length of $300$ km.}.

Correspondingly, the E2E packet loss probability is due to the backhaul failure, packet dropping in the ground BS's queue with a probability of $\varepsilon_{\rm{q}}^{\rm{g}}$, and decoding error of the G2A link with a probability of $\overline{\varepsilon}_{\rm{t}}^{\rm{ga}}$. Actually,  the reliability requirement in URLLC can be satisfied if the overall packet loss probability does not exceed $\varepsilon^{\rm{th}}$. Thus, the reliability can be guaranteed if
\begin{equation}
    \label{eq.reliabG2A}
     1-(1-\varepsilon_{\rm{b}})(1-\varepsilon_{\rm{q}}^{\rm{g}})(1-\overline{\varepsilon}_{\rm{t}}^{\rm{ga}})\leq \varepsilon^{\rm{th}}.   
\end{equation}
$\varepsilon_{\rm{b}}$ is the failure probability of backhaul link, which is modeled by Bernoulli process, and $1-\varepsilon^{\rm{th}}$ is the required reliability.

By using frequency diversity, a single BS can serve the AV with multiple links operating at different frequencies, thus it can increase the reliability based on the following equation
\vspace{-1mm}
\begin{equation}
    \label{eq.reliabDiver}
    \vspace{-1mm}
     1-(1-\varepsilon_{\rm{b}})(1-\varepsilon_{\rm{q}}^{\rm{g}})(1-\prod_{k=1}^{K}\overline{\varepsilon}_{\rm{t}}^{\rm{ga}})\leq \varepsilon^{\rm{th}},   
\end{equation}
where $K$ is the order of frequency diversity. The E2E delay can be decreased since the transmission delay of the G2A link is the minimum delay of $K$ independent transmission:
\begin{equation}
\label{eq.delayDiver}
     D_{\mathrm{b}} + D_{\mathrm{q}}^{\rm{g}} +  \min _{k} \left\{\overline{D}_{\mathrm{t}}^{\rm{ga}}\right\} \leq D_{\mathrm{max}}.
\end{equation}

\subsubsection{E2E Path through A2A Communication}
For the scenario of deploying an AV as a relay to transmit data to the AV of interest, the packet in addition to the DA2G communication path goes across relay AV's queue, with a delay of $D_{\mathrm{q}}^{\rm{a}}$, and A2A link, with an average delay of $\overline{D}_{\mathrm{t}}^{\rm{aa}}$. Hence, the delay components should satisfy
\begin{equation}
\label{eq.delayA2A}
     D_{\mathrm{b}} + D_{\mathrm{q}}^{\rm{g}} + \overline{D}_{\mathrm{t}}^{\rm{ga}} +  D_{\mathrm{q}}^{\rm{a}} + \overline{D}_{\mathrm{t}}^{\rm{aa}} \leq D_{\mathrm{max}}.
\end{equation}

Correspondingly, the reliability of the A2A communication path can be ensured if
\begin{equation}
\vspace{-2mm}
\begin{split}
    \label{eq.reliabA2A}
     1-(1-\varepsilon_{\rm{b}})(1-\varepsilon_{\rm{q}}^{\rm{g}})(1-\overline{\varepsilon}_{\rm{t}}^{\rm{ga}})(1-\varepsilon_{\rm{q}}^{\rm{a}})(1-\overline{\varepsilon}_{\rm{t}}^{\rm{aa}})\leq \varepsilon^{\rm{th}}.    
\end{split}
\end{equation}

If we consider a swarm of $M$ parallel coordinated AVs with single-hop transmission to serve the desired AV with joint decoding strategy, the E2E error probability and delay can be calculated by \eqref{eq.error.tot} and \eqref{eq.delay.tot}, respectively. In fact, it helps increase reliability by exploiting path diversity in the A2A link.

\subsubsection{E2E Path through HAP Communication}
For HAP, long distances of G2H and H2A links cause propagation delay in addition to previous delay components. Therefore, the E2E delay requirement of HAP is satisfied if 
\begin{equation}
\label{eq.delayH2A}
     D_{\mathrm{b}} + D_{\mathrm{q}}^{\rm{g}} + \overline{D}_{\mathrm{t}}^{\rm{gh}} + D_{\mathrm{p}}^{\rm{gh}} + D_{\mathrm{q}}^{\rm{h}} + \overline{D}_{\mathrm{t}}^{\rm{ha}} + D_{\mathrm{p}}^{\rm{ha}} \leq D_{\mathrm{max}},
\end{equation}
where $D_{\mathrm{p}}^{\rm{gh}}$ and $D_{\mathrm{p}}^{\rm{ha}}$ are the propagation delay of the G2H link and the H2A link, respectively.

The E2E packet loss probability of the HAP communication, similar to the A2A communication, can be computed as
\begin{equation}
\vspace{-2mm}
\begin{split}
    \label{eq.reliabH2A}
     1-(1-\varepsilon_{\rm{b}})(1-\varepsilon_{\rm{q}}^{\rm{g}})(1-\overline{\varepsilon}_{\rm{t}}^{\rm{gh}})(1-\varepsilon_{\rm{q}}^{\rm{h}})(1-\overline{\varepsilon}_{\rm{t}}^{\rm{ha}})\leq \varepsilon^{\rm{th}}.    
\end{split}
\end{equation}

\section{Numerical Results}\label{sec.res}

In this section, we evaluate the performance of different E2E connectivity paths consisting of different RATs and investigate how MC can ensure the stringent requirements of remote piloting the eVTOLs. We consider an urban scenario with macro cells. The system parameters are listed in Table \ref{tab.setup}, unless otherwise stated. The resource blocks (RBs) assigned to each AV consist of 4 RBs with 2 consecutive subcarriers. The subcarrier spacing is $0.2$ MHz. The allocated bandwidth of the X2Y link, $B^{\rm{xy}}$, $\rm{xy} \in \{\rm{ga},\rm{aa},\rm{ha}\}$, to transmit a packet is $0.4$ MHz\footnote{According to \cite{RA_urllc}, in URLLC, since the packet size is small the required bandwidth to transmit a packet does not exceed the coherence bandwidth which is around $0.5$ MHz.}, and frequency diversity is exploited as well for G2A link to increase the reliability. The dedicated bandwidth of the G2H link, $B^{\rm{gh}}$, is assumed to be $0.5$ MHz.

\begin{table}[!htb]
\centering
\caption{System Parameters \cite{3GPP38.811,U2U_comm,joint_UlDl}.}
\label{tab.setup}
\vspace{-0.2cm}
\begin{center}
\begin{tabular}{ |l|l| } 
 \hline
 Required packet loss probability $\varepsilon^{\rm{th}}$ & $10^{-5}$ \\
 \hline
 Delay threshold $D_{\rm{max}}$ & $10$ ms \& $50$ ms \\
 \hline
 Packet size $b$ & $32$ bytes \\ 
 \hline
 Average packet arrival rate of HAP $\lambda_{\rm{h}}$ & $10000$ packets/s \\ 
 \hline
 Average packet arrival rate of gBS $\lambda_{\rm{g}}$ & $1000$ packets/s \\ 
 \hline
 Average packet arrival rate of AV $\lambda_{\rm{a}}$ & $100$ packets/s \\ 
 \hline
 Queueing delay bound $D_{\max}^{\rm{q}}$ & $0.3$ ms \\ 
 \hline
 Queueing delay violation probability $\varepsilon_{\rm{q}}^{\rm{x}}$ & $10^{-7}$ \\ 
 \hline
 Backhaul failure probability $\varepsilon_{\rm{b}}$ & $10^{-6}$ \& $10^{-5}$ \\
 \hline
 Carrier frequency $f_{\rm{c}}$ & $2$ GHz \\
 \hline
 AV (gBS/HAP) Tx power & $23$ ($46$) dBm \\
  \hline
 AV Tx/Rx antenna gain $g_{\rm{a}}$ & $0$ dBi \\
  \hline
 Maximum gain of gBS antenna element $g_{\rm{E}}^{\max}$ & $8$ dBi \\
  \hline
 Maximum gain of HAP Tx/Rx antenna $g_{\rm{h}}^{\max}$ & $32$ dBi \\
  \hline
 AV (HAP) Rx noise figure & $9$ ($5$) dB \\
  \hline
 Number of gBS antenna elements $N$ & $8$ \\
  \hline
 Downtilt angle $\phi_{\rm{t}}$ & $102^{\circ}$ \\
  \hline
 Inter-Site Distance (ISD) & $500$ m \\
  \hline
 Height of gBS $h_{\rm{g}}$ & $25$ m \\
  \hline
 Altitude of AV $h_{\rm{a}}$ (HAP $h_{\rm{h}}$) & $300$ m ($20$ km) \\
  \hline
 Rice factor of G2A link $K_{\rm{ga}}$ & $5\sim12$ dB \\
  \hline
 Rice factor of A2A link $K_{\rm{aa}}$ & $12$ dB \\
  \hline
 Rice factor of G2H link $K_{\rm{gh}}$ & $5\sim15$ dB \\
  \hline
 Rice factor of H2A link $K_{\rm{ha}}$ & $12\sim15$ dB \\
  \hline
 Noise spectral density $N_0$ & $-174$ dBm/Hz \\
  \hline
 LoS (NLoS) Shadow fading standard deviation  & $4$ ($6$) dB \\
  \hline
\end{tabular}
\end{center}
\vspace{-0.75cm}
\end{table}

\begin{figure*}[h!]
	\begin{center}	
		\subfigure[]{
			\includegraphics[width=2.2in,trim={0.4cm 0cm 1cm 0.5cm},clip]{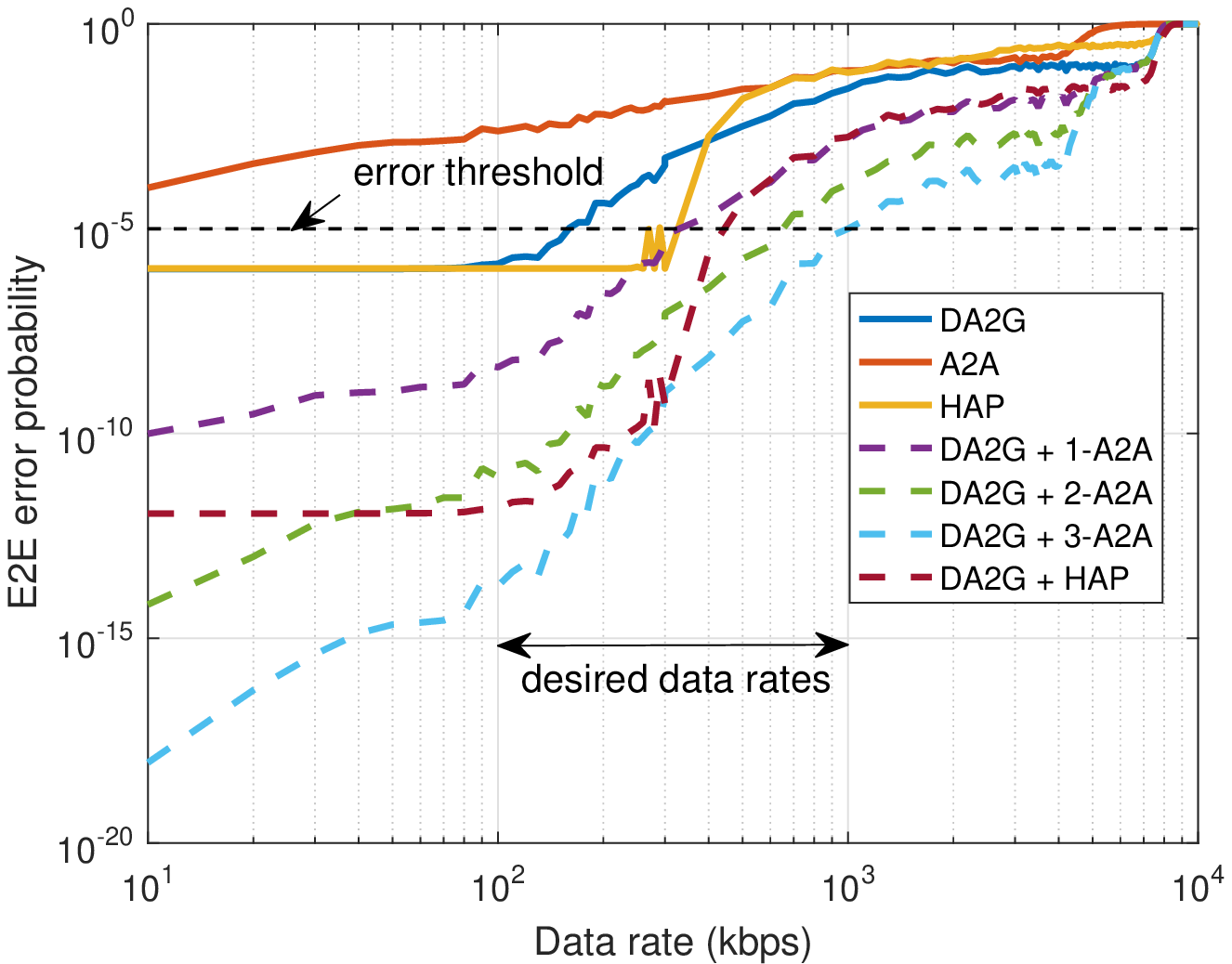} }
		\subfigure[]{
			\includegraphics[width=2.2in,trim={0.5cm 0cm 1cm 0.5cm},clip]{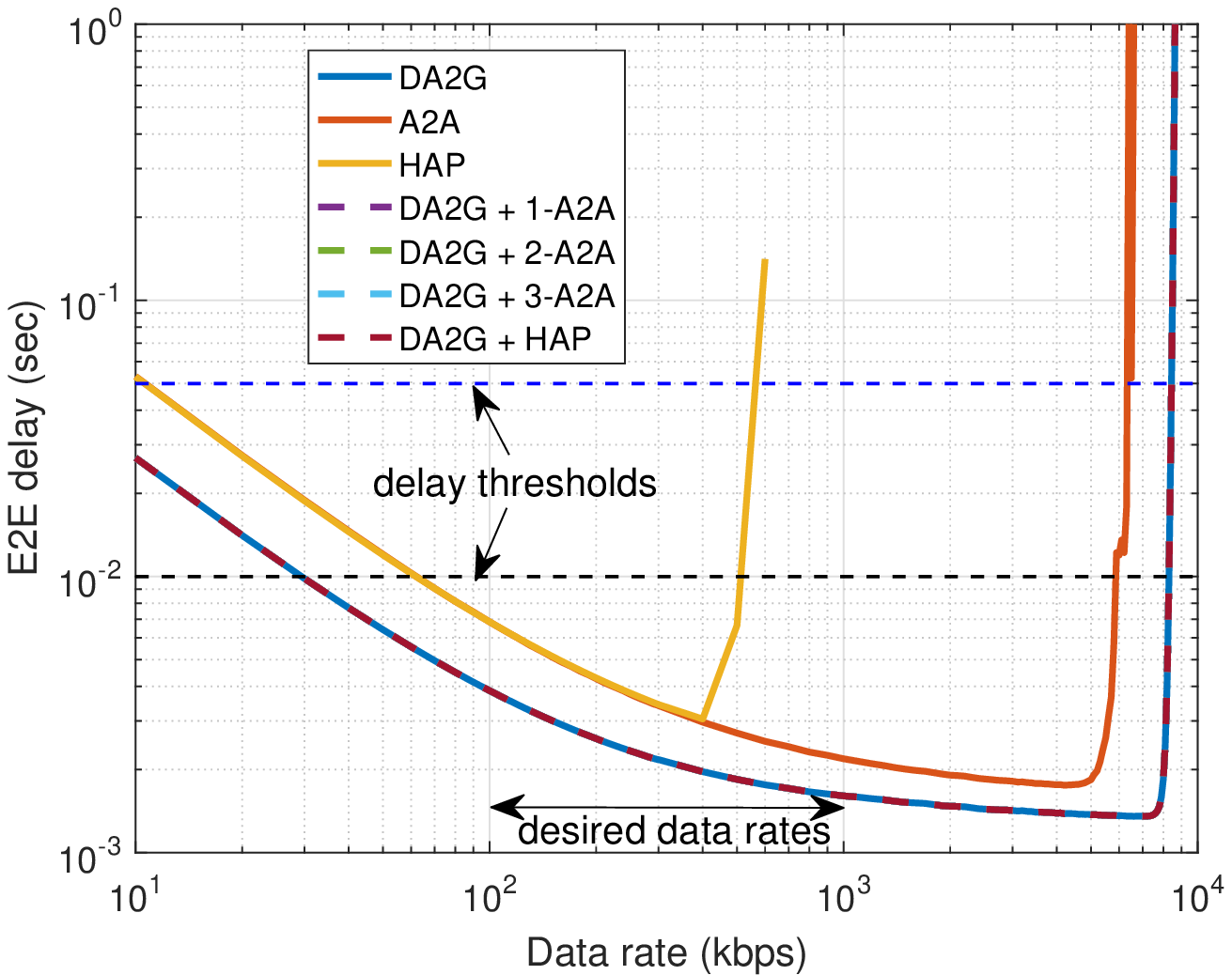} }
		\subfigure[]{
			\includegraphics[width=2.2in,trim={0.48cm 0cm 1cm 0.5cm},clip]{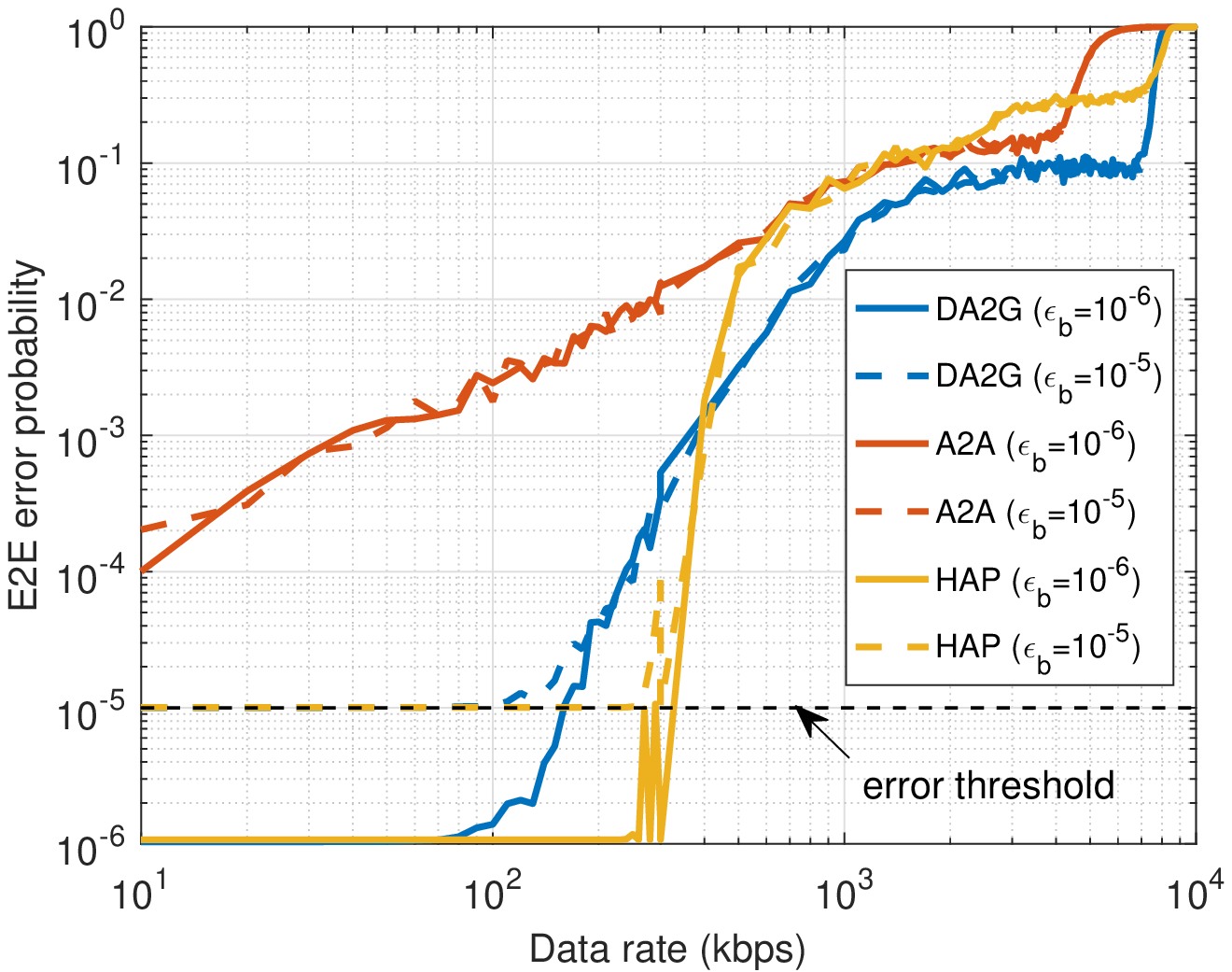} }

	\end{center}
	\vspace{-0.5cm}
	\caption{(a) E2E packet error probability and (b) E2E delay of different RATs and multi-path vs. data rate with $\varepsilon_{\rm{b}}=10^{-6}$, and $P_{\rm{interf}}=0.01$. (c) E2E error probability of different RATs vs. data rate with $P_{\rm{interf}}=0.01$ and different backhaul failure probabilities.}
	\label{fig.rate}
	\vspace{-0.3cm}
\end{figure*}

For all the RATs, we assume particular frequency band with full frequency reuse such that each link incurs probabilistic interference from all the corresponding links. In \cite{Rice_factor}, the Rician $K$-factor was found to increase exponentially with elevation angle between two nodes. Here for simplicity, we assume that the Rician factor of each link increases linearly with the elevation angle. The elevation angles are considered from $0^{\circ}$ to $90^{\circ}$ with a $10^{\circ}$ step, and the Rice factor is assumed to be constant in each interval.

We consider a hexagonal grid for the cellular terrestrial network consist of $3$ tiers, i.e., $37$ cells. $10$ AVs are located with uniform distribution at a fixed altitude over the considered cells, and $6$ of them are interfering with the AV of interest. The location of the desired AV's serving BS and the HAP projection on the ground is assumed at the origin. The horizontal distance of the HAP and its ground station is set as $5$ km.

Fig. \ref{fig.rate} shows the E2E  error probability and delay of different paths and their combinations with respect to the data rate when an AV's 2D distance from the serving ground BS, $r_{\rm{ga}}$, is $150$ m. The probability of backhaul failure, $\varepsilon_{\rm{b}}$, and the probability of interference, $P_{\rm{interf}}$, is considered as $10^{-6}$ and $0.01$, respectively. As shown in Fig. \ref{fig.rate}(a) and Fig. \ref{fig.rate}(b), E2E path with A2A link can not individually achieve the reliability, while E2E path with DA2G communication guarantees the QoS requirements with $10$ ($50$) ms delay threshold from $30$ (less than $10$) kbps up to $200$ kbps. Similarly, E2E path with HAP guarantees the requirements with $10$ ($50$) ms delay threshold from $60$ ($10$) kbps up to $300$ kbps. It is observed that DA2G with one A2A link is an alternative for reliable communication in the desired range of data rates up to $300$ kbps for both delay thresholds. Combining DA2G path with two (three) A2A parallel paths can satisfy the reliability up to $600$ kbps ($1$ Mbps). Combining DA2G and HAP paths can satisfy the requirements up to $400$ kbps.

\subsection{Impact of Backhaul Failure Probability on Reliability}
Previous analysis in Fig. \ref{fig.rate}(a) and Fig. \ref{fig.rate}(b) was based on the backhaul failure probability of $10^{-6}$ which is lower than $\varepsilon^{\rm{th}}=10^{-5}$. This is an optimistic assumption on the backhaul link. Now, we assume that probability of backhaul failure is $10^{-5}$, i.e., backhaul link is the bottleneck. Hence, transmitting signal through one path will never fulfill the reliability. Fig. \ref{fig.rate}(c) illustrates the E2E error probability of different RATs for the two considered backhaul failure probabilities. It is observed that the overall error is affected by the backhaul failure in low data rates, where the decoding error is low, and the backhaul error is dominating. However, in higher data rates more than $400$ kbps, where the decoding error is high, and the backhaul failure is negligible, the performance are the same. Therefore, MC is a solution to ensure the QoS requirements.

\subsection{Effect of Interference}
\begin{figure}[!htb]
    \centering
            \includegraphics[width=0.99\columnwidth]{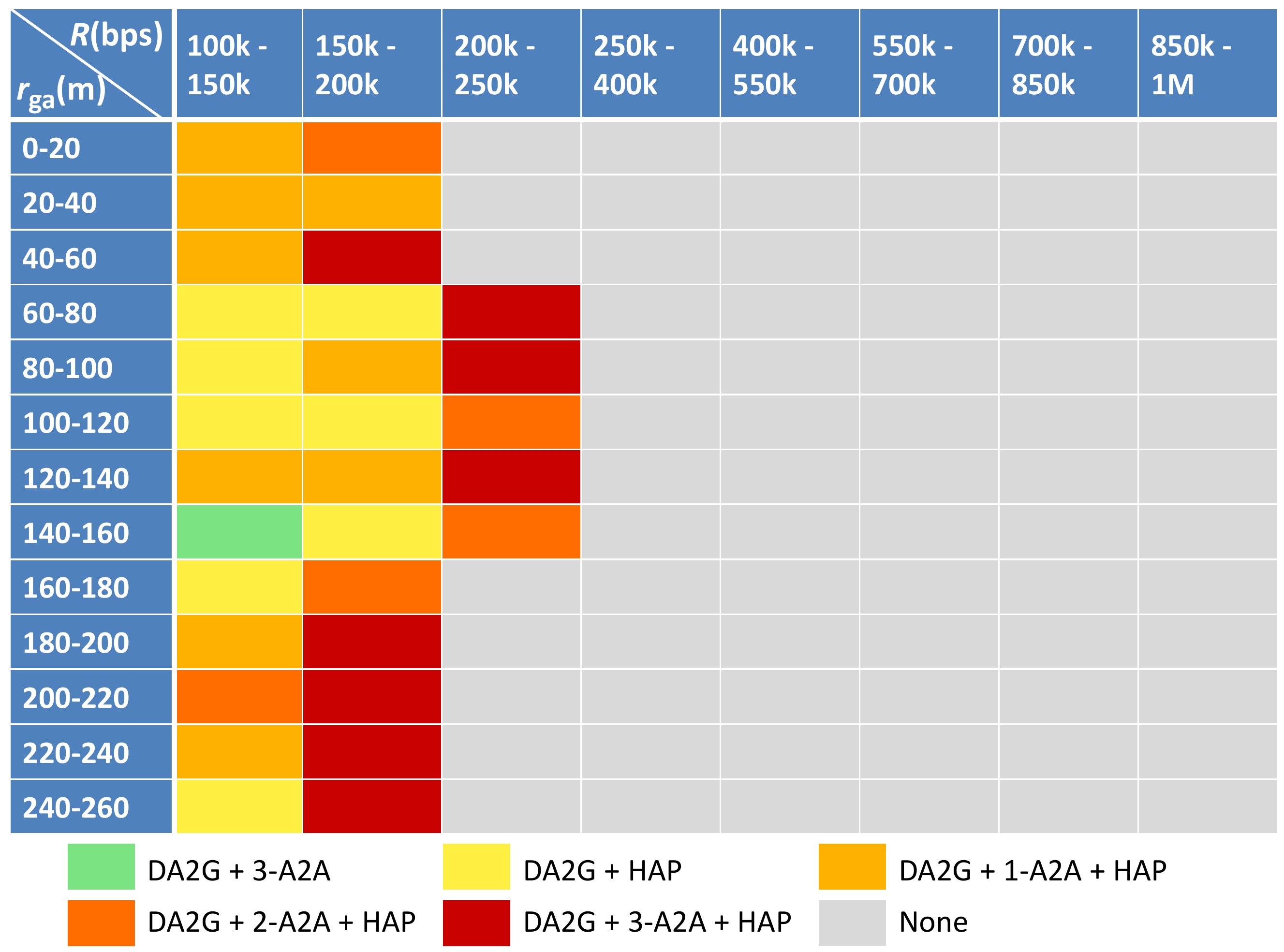}
            \caption{Operating region of different multi-path to guarantee the QoS of eVTOL with  $\varepsilon_{\rm{b}}=10^{-5}$ and $P_{\rm{interf}}=0.1$.} 
            \label{fig.table1}
            \vspace{-0.5cm}
\end{figure}

We achieve the average E2E reliability and delay for various 2D distances between the AV and its serving BS in the cell range with $\varepsilon_{\rm{b}}=10^{-5}$ and $P_{\rm{interf}}=0.1$. Fig. \ref{fig.table1} indicates the minimum required number of paths based on an order with a descending preference as DA2G, A2A, and HAP communication with atmost $3$ A2A links, in each region to satisfy the QoS requirements. Each region in Fig. \ref{fig.table1} shows the desired data rate,  $R$, and the 2D distance of the AV and its serving BS, $r_\mathrm{ga}$. It is observed that when $r_\mathrm{ga}$ is within $0-20$ m, three E2E paths consisting of DA2G, A2A, and HAP are required to support data rates of $100-150$ kbps, which is denoted as ``DA2G + 1-A2A + HAP". While it demands one more path with another A2A relay link, i.e., ``DA2G + 2-A2A + HAP" to satisfy the requirements from $150-200$ kbps. Data rates up to $250$ kbps are provided when the AV's 2D distance is $60-160$ m with combination of DA2G, two/three A2A paths, and HAP. It is observed that higher data rates can be met by none of the path combinations. The reason is the strong interference on the radio links due to the LoS conditions, especially DA2G with the neighboring BSs, which increases the decoding error and leads to severe degradation of the reliability and latency performance.

\begin{figure}[!htb]
    \centering
            \includegraphics[width=0.99\columnwidth]{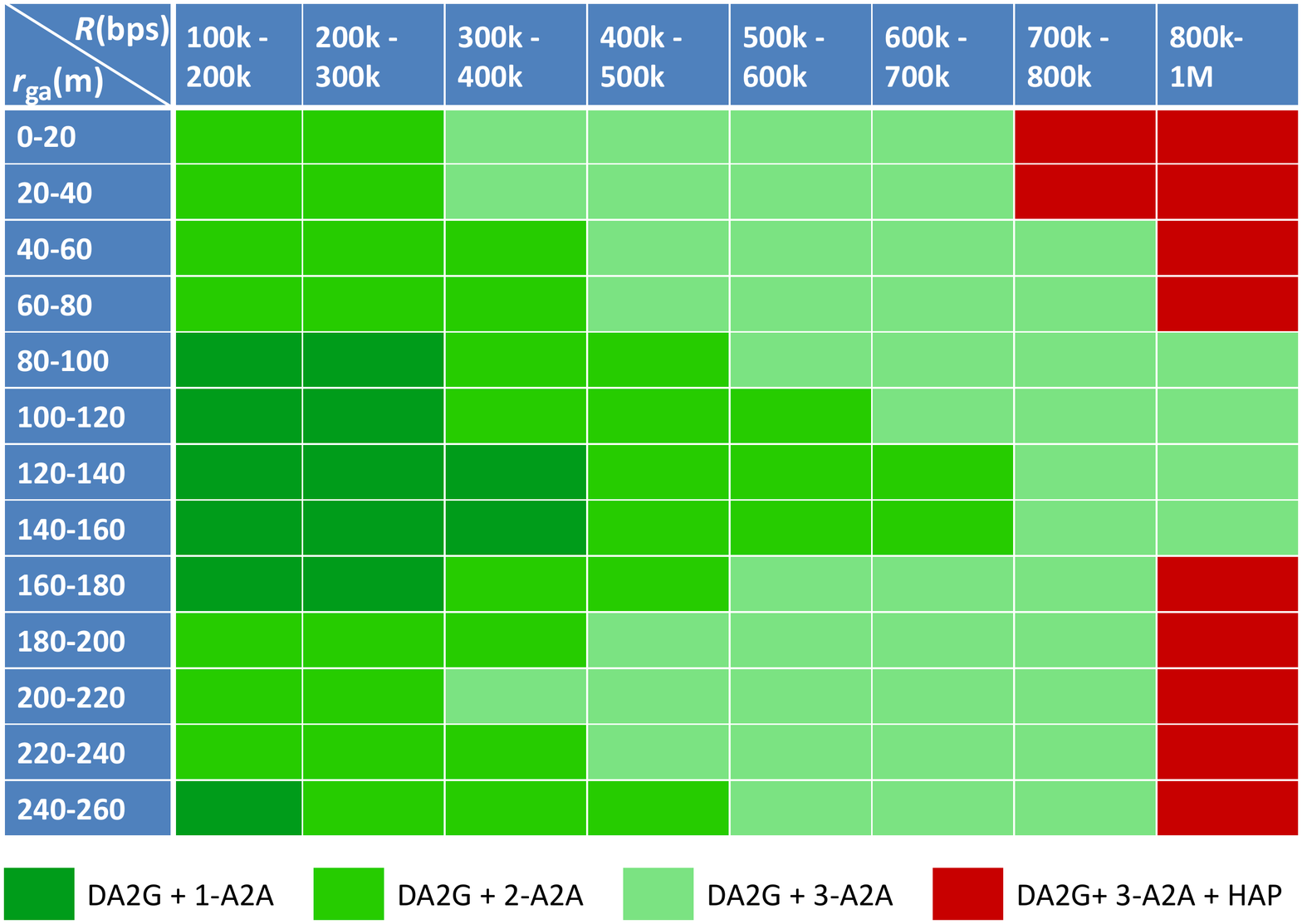}
            \caption{Operating region of different multi-path to guarantee the QoS of eVTOL with  $\varepsilon_{\rm{b}}=10^{-5}$ and $P_{\rm{interf}}=0.01$.} 
            \label{fig.table2}
            \vspace{-0.3cm}
\end{figure}

Fig. \ref{fig.table2} shows the achieved rates with the use of different multi-paths when $P_{\rm{interf}}=0.01$.
In this case, all the desired data rates from $0.1-1$ Mbps can be satisfied by the RAT diversity. The QoS requirements in most of the regions can be guaranteed by DA2G and multi A2A paths. Actually, the achieved graph confirms the radiation pattern of the ground BS antenna, with the most directivity of side lobes in elevation angle of $63^{\circ}$. High data rates when the AV flies in the cell center/edge, with ISD of $500$ m, require the HAP connection, too. Overall, it indicates that even with highly efficient interference mitigation, a single RAT will not be able to provide seamless connectivity for all the desired data rates, and MC is needed for remote piloting of AVs.
\vspace{-0.2cm}

\section{Conclusion}\label{sec.con}
In this paper, we have studied the downlink communication of an AV in finite blocklength regime with MC under practical antenna configurations. To this end, we have integrated multiple RATs including DA2G, A2A, and HAP communications. E2E reliability and latency are characterized as functions of the system parameters such as required data rate, 2D distance between the AV and its serving ground BS, probability of interference, and backhaul failure probability. We have shown that the overall performance of different links under practical antenna settings is highly limited due to the LoS interference. The empirical results reveal that when there is high interference on the wireless links, safe operation of a special type of AVs, i.e, eVTOLs, even with HAP communication in addition to the DA2G and A2A paths, can be provided up to the data rate of $250$ kbps.
Moreover, we have demonstrated that even with interference mitigation, regardless of the backhaul link as a bottleneck, MC is a key enabler for the service requirements by DA2G and A2A links in most of the cases, while HAP is required as an auxiliary link for data rates higher than $700$ kbps on the cell center and the cell edge in the evaluated scenario.

\section*{Acknowledgment}
This work was supported by EU Celtic Next Project, 6G for Connected Sky (6G-SKY).

\bibliographystyle{IEEEtran}
\bibliography{paper.bib}

\end{document}